\documentclass[journal]{IEEEtran}
\usepackage{graphicx}
\usepackage{indentfirst}
\usepackage{inputenc}
\usepackage{multicol}
\usepackage[switch,mathlines]{lineno}
\usepackage{soul}
\usepackage{cite}
\usepackage{CJK}
\usepackage[cmex10]{amsmath}
\usepackage{algorithmic}
\usepackage{array}
\usepackage{bm}
\usepackage{amssymb}
\usepackage{color}
\usepackage{nomencl}
\usepackage{pdfpages}
\usepackage{booktabs} 
\usepackage{multirow}
\usepackage{threeparttable}
\usepackage[ruled, linesnumbered]{algorithm2e}
\usepackage[colorlinks, linkcolor=blue, citecolor=blue, urlcolor=blue]{hyperref}
\usepackage[colorlinks, citecolor=blue]{hyperref}
\ifCLASSOPTIONcompsoc
 \usepackage[caption=false,font=normalsize,labelfont=sf,textfont=sf]{subfig}
\else
 \usepackage[caption=false,font=footnotesize]{subfig}
\fi
\usepackage{url}
\begin{document}

\title{Neural Network Certification Informed Power System Transient Stability Preventive Control with Renewable Energy}

\author{{Tong Su, \IEEEmembership{Student Member, IEEE},
    Junbo Zhao, \IEEEmembership{Senior Member,~IEEE}
    }

\thanks{This work was supported by the U.S. Department of Energy Office of Science and Office and Electricity Advanced Grid Modeling program.
\par T. Su and J. Zhao are with the Department of Electrical and Computer Engineering, University of Connecticut, Storrs, CT 06269, USA (e-mail: tongsu@uconn.edu; junbo@uconn.edu).
}}
\markboth{}%
{Shell \MakeLowercase{\textit{et al.}}: Bare Demo of IEEEtran.cls for Journals}

\maketitle
\begin{abstract}
Existing machine learning-based surrogate modeling methods for transient stability constrained-optimal power flow (TSC-OPF) lack certifications in the presence of unseen disturbances or uncertainties. This may lead to divergence of TSC-OPF or insecure control strategies. This paper proposes a neural network certification-informed power system transient stability preventive control method considering the impacts of various uncertainty resources, such as errors from measurements, fluctuations in renewable energy sources (RESs) and loads, etc. A deep belief network (DBN) is trained to estimate the transient stability, replacing the time-consuming time-domain simulation-based calculations. Then, DBN is embedded into the iterations of the primal-dual interior-point method to solve TSC-OPF. To guarantee the robustness of the solutions, the neural network verifier $\alpha, \beta$-CROWN to deal with uncertainties from RESs and loads is proposed. The yielded certification results allow us to further adjust the transient stability safety margin under the iterated TSC-OPF solution process, balancing system security and economics. Numerical results on a modified western South Carolina 500-bus system demonstrate that the proposed method can efficiently and quickly obtain the safety-verified preventive control strategy through RES curtailment and generator dispatch with only a slight increase in cost.
\end{abstract}


\section{Introduction}\label{sec1}
\IEEEPARstart{N}{eural} networks (NNs) have achieved great success in various power system applications, such as renewable energy source (RES) forecasting \cite{wang2019review}, load forecasting \cite{almalaq2017review}, and stability estimation \cite{alimi2020review}. However, the reliability and robustness of NNs have raised concerns, as they are vulnerable to adversarial examples, meaning that imperceptible perturbations of test samples might unexpectedly change the NNs estimations \cite{szegedy2013intriguing, biggio2018wild, fawzi2017robustness}. 

For practical deployments of NNs in power systems, the inputs to NNs are typically measurements or forecasts. The measurement errors associated with different measurement devices can affect the estimations made by NNs \cite{ieee2011std, zhao2017impact}. Moreover, measurement data may be subjected to false data injections and cyber-attacks \cite{liang2016review, manandhar2014detection}. The adversarial samples generated in this way can be very close to true values but may lead to opposite NN estimations, such as misjudging an unstable scenario as stable, failing to trigger necessary controls, leading to system failures, blackouts, or even more severe accidents \cite{yuan2019adversarial, ren2022robustness}. In addition, the forecasts of RESs or loads have high uncertainty, and control strategies based on deterministic forecasts may be unsafe when executed. Therefore, robust estimation and control considering all perturbations are crucial.

There has been some research on the robustness analysis of NNs, which can be categorized into adversarial sample generation, adversarial samples-based NNs' robustness enhancement, and verification of NNs' robustness. The core idea for generating adversarial samples is similar, which is to set neighborhoods with different norms based on the samples that need to be verified. The size of the neighborhood depends on the perturbation magnitude of the sample point. The adversarial samples are situated throughout the entire neighborhood, and the size of this neighborhood is determined by the perturbation amplitude of the samples. Then, these adversarial samples are used to retrain the NNs, enhancing their robustness. For example, \cite{tian2021joint} proposes a state-perturbation-based adversarial example and the false data injection attacks method can carry out attacks stealthy to both conventional bad data detectors and deep learning-based detectors. In \cite{tian2021adversarial}, a signal-specific method and a universal signal-agnostic method are proposed, which can respectively generate perturbations to misclassify most natural signals with high probability and have a higher transfer rate of black-box attacks. In addition, adversarial training is adopted to defend systems against adversarial attacks in \cite{tian2021adversarial}. 
Robustness verification is divided into two types, one is based on Monte Carlo sampling (MCS) to find the largest neighborhood, where the sample being verified will not be misclassified. The larger neighborhood range corresponds to the higher robustness of the test sample. In \cite{ren2022robustness}, the MCS-based method is applied to calculate the Lipschitz constant within a norm ball to approximate the robust index of NNs. Similarly, \cite{ren2022universal} employs a randomized smoothing algorithm to transform any classifier, which performs well under Gaussian noise, into a new classifier that is certifiably robust against adversarial examples. Given that it is not feasible to precisely calculate the classification probabilities of the base classifier, and it is also challenging to accurately assess the prediction and robustness ability of the smoothed classifier, the MCS is employed for both tasks, ensuring success with arbitrarily high probability. Then, based on sampling or probabilistic methods, it's impossible to guarantee with 100\% certainty that all points within the neighborhood are reliably classified, leaving a risk of misclassification. Another category is analytical methods, for example, \cite{ren2021vulnerability} calculates the norm distance from samples to the separating affine hyperplane as an index of robustness through the linearization of the model. However, for non-linear and non-convex models, achieving the optimal solution is challenging regardless of linearization. There may exist smaller distances to the separating affine hyperplane that lead to misclassification. \cite{venzke2020verification} uses mixed-integer programming (MIP) to compute the minimum distance to a sample which changes the classification, where the ReLU function is piecewise linearized through binary variables. While MIP has proven effective in verifying the robustness of NNs, particularly those employing piecewise linear activation functions such as ReLU, they encounter considerable obstacles when applied to NNs featuring more complex and general nonlinearities. Moreover, MIP is highly time-consuming for verifying large networks. 

To rigorously ensure that the robustness within the p-norm ball of the test example is fully verified, and applicable to general activation functions, the CROWN is proposed. CROWN is an efficient bound propagation-based verification algorithm that backpropagates a linear inequality through the network, relaxing activation functions with linear bounds. Additionally, by adaptively selecting the linear approximation when computing certified lower bounds of minimum adversarial distortion, it can improve the certified lower bound \cite{zhang2018efficient}. $\alpha$-CROWN enhances the CROWN verifier by optimizing both intermediate and final layer bounds using variable $\alpha$, offering superior effectiveness over linear programming (LP) by more efficiently tightening intermediate layer bounds \cite{xu2020fast}. Then, $\beta$-CROWN extends the CROWN verifier by integrating ReLU split constraints in branch and bound (BaB) into the bound propagation process to optimize parameter $\beta$ \cite{wang2021beta}.  \cite{shi2024neural} extends BaB-based verification for non-ReLU and general nonlinear functions, achieving significant improvements in verifying NNs with non-ReLU activation functions, such as Transformer and long short-term memory networks. 
Furthermore, GPUs can effectively parallelize and accelerate bound propagation in the BaB process for $\beta$-CROWN and GCP-CROWN \cite{xu2020automatic}.

Preventive control aims to prepare the system for potential credible contingencies through generator dispatch and RES curtailment \cite{yuan2021robustly}. Transient stability constrained-optimal power flow (TSC-OPF) is widely used to find the optimal operating point under transient stability constraints, but it relies on time-consuming time-domain simulations (TDSs), making it challenging for online applications in large-scale power systems. Several alternatives to TDSs are proposed to accelerate the calculation, such as transient energy function method \cite{bhui2016real}, extended equal-area criterion \cite{yuan2020preventive}, trajectory sensitivity method \cite{nguyen2003dynamic}, etc. Deep learning methods have been applied as surrogate models for TDSs to estimate power system transient stability, achieving significant acceleration from several to tens of seconds to less than 0.01 seconds \cite{su2023deep}. Due to measurement or forecast errors, as well as cyber-attacks, the robustness of control strategies is critical. However, there is currently no fast and complete NN verification method for high-dimensional, nonlinear, and non-convex transient stability control problems.

This paper proposes an NN robustness certification informed power system transient stability preventive control method. The main contributions are summarized as follows:
\begin{enumerate}
\item{Deep belief networks (DBNs) are used to replace time-consuming TDSs in power systems transient stability estimation, significantly accelerating this process. Additionally, DBNs are analytically integrated into the primal-dual interior-point method (PDIPM), serving as surrogate models to facilitate solving TSC-OPF.}
\item{Due to measurement or forecast errors, as well as cyber-attacks, the uncertainty intervals of RESs and loads are constrained within a norm ball. 
To guarantee the robustness of the control solutions, the neural network verifier $\alpha, \beta$-CROWN is developed to address uncertainties from RESs and loads. The certification results allow further adjustment of the transient stability safety margin by iterative TSC-OPF process. The proposed method can effectively balance system security and economics while being scalable to large-scale systems. Based on our knowledge, this is also the first work on NN robustness certification-based TSC-OPF.}
\end{enumerate}

The rest of the paper is organized as follows. Section \ref{sec2} presents the TSC-OPF problem. Section \ref{sec3} introduces the proposed NN robustness verification-based preventive control framework. Results are presented and analyzed in Section \ref{sec4}, and finally, Section \ref{sec5} concludes the paper.

\section{Problem Statement}\label{sec2}
TSC-OPF is commonly applied to determine the optimal operating point while considering static and transient stability constraints, and its mathematical formulation is as follows \cite{zimmerman2010matpower}

\subsection{Objective Function}\label{sec2a}
The objective function is to minimize the cost of generator (SG) dispatch and RES curtailment.
\begin{equation}
\label{Eq_objective}
    \operatorname{min} \sum_{i \in \mathcal{G}}\left(a P_{\text{SG}_i, t}^{2}+b P_{\text{SG}_i, t}+c\right)+\sum_{i \in \mathcal{R}} d \left(\overline{P}_{\text{IBR}_i, t} - P_{\text{IBR}_i, t}\right)
\end{equation}
where $t$ is a time instant; $P_{\text{SG}}$ and $P_{\text{IBR}}$ denote the active power generations of SGs and RESs/inverter-based resources (IBRs), respectively; $\overline{P}_{\text{IBR}}$ is the forecasted maximum value of $P_{\text{IBR}}$; $\mathcal{G}$ and $\mathcal{R}$ denote the set of $P_{\text{SG}}$ and $P_{\text{IBR}}$, respectively; $a$, $b$, and $c$ are fuel cost coefficients of $P_{\text{SG}}$; $d$ is the curtailment cost coefficient of $P_{\text{IBR}}$.

\subsection{Power Flow Equations}\label{sec2b}
The active power balance equation $\bm g_P$ and reactive power balance equation $\bm g_Q$ are
\begin{equation}
\label{Eq_PF_P}
    \bm{g}_{P}\left(\bm{\Theta}, \bm{V}_{m}, \bm{P}_{g}\right) =\bm{P}_{\text {bus }}\left(\bm{\Theta}, \bm{V}_{m}\right)+\bm{P}_{d}-\bm{C}_{g} \bm{P}_{g}=\bm 0
\end{equation}
\begin{equation}
\label{Eq_PF_Q}
    \bm{g}_{Q}\left(\bm{\Theta}, \bm{V}_{m}, \bm{Q}_{g}\right) =\bm{Q}_{\text {bus }}\left(\bm{\Theta}, \bm{V}_{m}\right)+\bm{Q}_{d}-\bm{C}_{g} \bm{Q}_{g}=\bm 0
\end{equation}
where $\bm P_{\text {bus}}$ and $\bm Q_{\text {bus}}$ denote the bus active and reactive power injection vectors, respectively; $\bm{\Theta}$ and $\bm{V}_m$ denote the bus voltage angle and magnitude vectors, respectively; $\bm P_g$ and $\bm Q_g$ denote the active and reactive power injection vectors by SGs and IBRs, respectively; $\bm {P}_{d}$ and $\bm {Q}_{d}$ denote the active and reactive load demand vectors, respectively; $\bm{C}_g$ is the connection matrix of SGs and IBRs.

\subsection{Static Constraints}\label{sec2c}
Static constraints include line capacity constraints, bus voltage magnitude constraints, SG active and reactive power generation constraints, IBRs active and reactive power generation constraints.
\begin{equation}
\label{Eq_line_Constraint}
    \left|\bm P_{\text{Line}, t}\right| \leq \bm P_{\text{Line}}^{\max} ~~~~~~~~~~ \bm V_m^{\min} \leq \bm V_{m, t} \leq \bm V_m^{\max}
\end{equation}
\begin{equation}
\label{Eq_P_SG_Constraint}
    \bm P_\text{SG}^{\min} \leq \bm P_{\text{SG}, t} \leq \bm P_\text{SG}^{\max} ~~~~~ \bm Q_\text{SG}^{\min} \leq \bm Q_{\text{SG}, t} \leq \bm Q_\text{SG}^{\max}
\end{equation}
\begin{equation}
\label{Eq_PQ_IBR_Constraint}
    \bm 0 \leq \bm P_{\text{IBR}, t} \leq  \bm{\overline{P}}_{\text{IBR}, t} \leq \bm S_\text{IBR}^\text{rated} ~~~ \bm Q_\text{IBR}^{\min} \leq \bm Q_{\text{IBR}, t} \leq  \bm Q_\text{IBR}^{\max}
\end{equation}
\begin{equation}
\label{Eq_S_IBR}
    \bm Q_\text{IBR}^{\max/\min} = +/-\sqrt{(\bm S_\text{IBR}^\text{rated})^2 - (\bm P_\text{IBR})^2}
\end{equation}
where $\bm P_l$ and $\bm P_{l}^{\max}$ denote the line active power flow and its maximum value, respectively; $\bm V_m^{\max}$ and $\bm V_m^{\min}$ are upper and lower limits of $\bm V_m$; $\bm P_\text{SG}^{\max}$ and $\bm P_\text{SG}^{\min}$ are upper and lower limits of $\bm P_\text{SG}$; $\bm Q_\text{SG}^{\max}$ and $\bm Q_\text{SG}^{\min}$ are upper and lower limits of $\bm Q_\text{SG}$; $\bm Q_\text{IBR}^{\max}$ and $\bm Q_\text{IBR}^{\min}$ are upper and lower limits of $\bm Q_\text{IBR}$; $S_\text{IBR}^\text{rated}$ is the rated apparent power.

\subsection{Transient Stability Constraint}\label{sec2d}
The transient stability index (TSI) can reflect the system's stability and is calculated using a set of differential-algebraic equations (DAEs) of the power system.
\begin{equation}
\label{Eq_DAEs}
\begin{cases}
    \dot{\bm{x}}_t=\bm{f}(\bm{x}_t, \bm{y}_t, \bm{u}, \bm{p}, \bm{\tau})\\
    \bm{0}=\bm{g}(\bm{x}_t, \bm{y}_t, \bm{u}, \bm{p}, \bm{\tau})
\end{cases}
\end{equation}
\begin{equation}
\label{Eq_TSI}
    \operatorname{TSI}(\bm{x})=\frac{360^\circ-\delta^{\max}}{360^\circ+\delta^{\max}} \times 100 > \lambda \geq 0
\end{equation}
where $\bm{x}_t$ represents the state variables, such as the rotor angle $\delta$ and rotor speed $\omega$ of SGs, virtual rotor speeds and angles of IBRs, etc; $\bm{y}_t$ is the algebraic variables, such as voltage and current; $\bm{p}$ denotes the physical parameters \cite{kundur1994power}; $\bm{u}$ represents the control strategies; $\bm{\tau}$ denotes the uncertainties caused by measurement and forecast errors and fluctuations in RESs and loads; $\delta^{\max}$ denotes the maximum rotor angle difference of any two generators during the TDSs; $\lambda$ denotes the boundary between transient stability and instability. TSI is positively correlated with the system's transient stability, and it requires TSI to be greater than 0, meaning $\text{TSI} > \lambda \geq 0$ \cite{gupta2018online}. Due to uncertainties $\bm{\tau}$, the scenario during the execution of the preventive control strategy may deviate from the one used to generate the strategy, potentially leading to control failure. Therefore, robustness needs to be verified.

\section{Proposed NN Robustness Verification-Based Power System Preventive Control}\label{sec3}

\subsection{DBN-Based Transient Stability Surrogate Model}\label{sec3a}
DBN is a stack of multiple layers of restricted Boltzmann machines, followed by a fully connected layer. DBN can improve estimation accuracy through layer-by-layer pre-training of RBMs. The details of DBN are shown in \cite{su2021deep}. The input vector of DBN is $\bm{x} = \left[\bm{P}_{\text{IBR}}, \bm{P}_{\text{SG}}, \bm{P}_{d}, \bm{Q}_{d}\right]^{T}$. The output variable is the estimated TSI, i.e., $\widetilde{\text{TSI}}$. The DBN-based transient stability constraint is
\begin{equation}
\label{Eq_DBN_Estimation}
    \widetilde{\text{TSI}} = \mathcal{DBN}(\bm{\widetilde{x}}, \bm W , \bm b) > \lambda \geq 0
\end{equation}
where $\bm{\widetilde{x}}$ denotes the input of DBN, including control strategies $\bm{\widetilde{P}}_{\text{IBR}}$ and $\bm{\widetilde{P}}_{\text{SG}}$, as well as the forecasted $\bm{\widetilde{P}}_d$ and $\bm{\widetilde{Q}}_d$; $\bm W$ and $\bm b$ denote the weights and biases of DBN. By using \eqref{Eq_DBN_Estimation}, the speed of transient stability assessment can be improved by hundreds or even thousands of times.

\subsection{Projected Gradient Descent Attack}\label{sec3b}
The goal of a projected gradient descent (PGD) attack is to generate adversarial examples $\bm x_{\text{adv}}$ that are sufficiently close to the original input $\bm{\widetilde{x}}$ to attack the NN, aiming to find points in the input space that, with small perturbations, can lead the NN to output incorrect classification results. PGD achieves this by maximizing the network's loss function $\mathcal{L}(\bm{x}_{\text{adv}}, y)$ (where $y$ is the correct label, i.e., TDS-based TSI) at each iteration:
\begin{equation}
\label{Eq_PGD_Attack}
    \bm{x}_{\text{adv}}^{t+1} = \Pi_{\mathcal{C}} \left( 
    \bm{x}_{\text{adv}}^t + \eta \cdot \text{sign} \left( \nabla_{\bm{x}} \mathcal{L} \left( \bm{x}_{\text{adv}}^t, y \right) \right) \right)
\end{equation}
where $\mathcal{C}$ defines the perturbation range of the input $\bm x$. For power systems with element-wise bounds, we consider $\mathcal{C}$ as an $\ell_\infty$ ball around the forecasts $\bm{\widetilde{x}}$: $\mathcal{C} = \left\{\bm x \ \middle| \ \|\bm x - \bm{\widetilde{x}}\|_\infty \leq \bm \epsilon \right\}$; $\Pi$ is the projection operator, ensuring that the updated $\bm{x}_{\text{adv}}^t$ stays within the set $\mathcal{C}$; $\eta$ denotes the step size. PGD attacks are fast but cannot provide global robustness guarantees, only revealing the NN's vulnerability at certain points.

\subsection{Linear Programming Verifier}\label{sec3c}
\begin{figure}[htb]
  \centering
  \includegraphics[width=8.8cm]{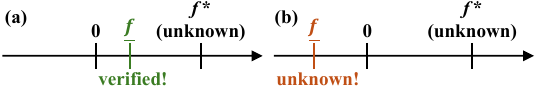}
  \caption{Incomplete verification. \textbf{(a)}: $\underline f \geq 0$, $\bm{\widetilde{x}}$ is verified. \textbf{(b)}: $\underline f < 0$, the verification result is unknown.}
  \label{Fig_incomplete}
\end{figure}

A $L$-layer NN function $f$ is defined as $f(\bm x) = z^{(L)}(\bm x)$, where $\bm z^{(i)}(\bm x) = \bm{W}^{(i)} \bm{\hat{z}}^{(i-1)}(\bm x) + \bm{b}^{(i)}$; $\bm{\hat{z}}^{(i)}(\bm x) = \sigma\left( \bm z^{(i)}(\bm x) \right)$; $\bm{\hat{z}}^{(0)}(\bm x) = \bm x$; $\sigma$ denotes the ReLU activation function; $\bm z^{(i)}$ and $\bm{\hat{z}}^{(i)}$ represent the pre-activation and post-activation values. The NN verification can be formulated as an optimization problem as follows \cite{wang2021beta}:
\begin{equation}
\label{Eq_NN_Optimization}
\begin{split}
    \min ~ & f(x):=z^{(L)}(\bm x) \\ \text { s.t. }  ~ & \bm z^{(i)} = \bm{W}^{(i)} \bm{\hat{z}}^{(i-1)}+\bm{b}^{(i)}, ~~ \bm{\hat{z}}^{(i)} = \sigma\left(\bm z^{(i)}\right), \\ & \bm x \in \mathcal{C}, ~~ i \in\{1, \cdots, L-1\}
\end{split}
\end{equation}
For the control strategy $\bm{\widetilde{x}}$, with $\widetilde{\text{TSI}} = f(\bm{\widetilde{x}}) > 0$, if it can be proven that $f(\bm x) > 0$, $\forall \bm x \in \mathcal{C}$, then $\bm{\widetilde{x}}$ is successfully verified. NN verification is categorized into complete and incomplete verification. Complete verifiers can solve \eqref{Eq_NN_Optimization} exactly, i.e., $f^* = \min f(\bm x)$, $\forall \bm x \in \mathcal{C}$. In contrast, incomplete verifiers typically relax the non-convexity of NNs to obtain a tractable lower bound of the solution, i.e., $\underline f \leq f^*$. If $\underline f \geq 0$, then $f^* \geq 0$ so $\bm{\widetilde{x}}$ is verified, as shown in Fig. \ref{Fig_incomplete}-a. However, if $\underline f < 0$, the verification result is unknown, as shown in Fig. \ref{Fig_incomplete}-b.

\begin{figure*}[htb]
  \centering
  \includegraphics[width=18.1cm]{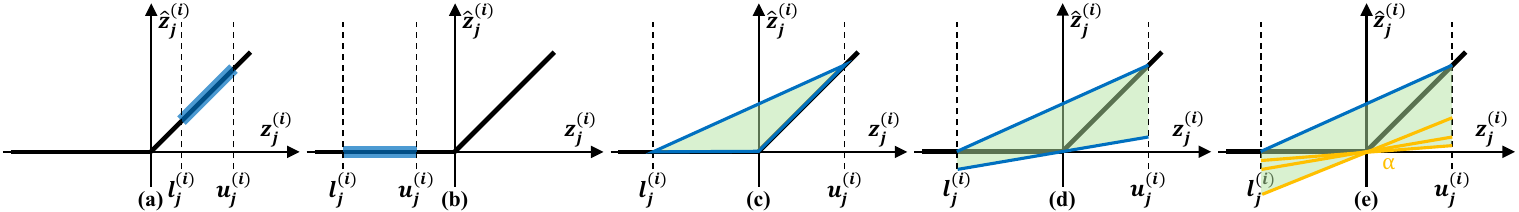}
  \caption{Relaxations of a ReLU. (a)(b) No relaxation when $l_j^{(i)} \geq 0$ or $l_j^{(i)} \leq 0$; (c) “triangle” relaxation when $l_j^{(i)} \leq 0 \leq u_j^{(i)}$ in LP verifiers; (d) linear relaxation in CROWN with fixed lower bounds; (e) linear relaxation in $\alpha$-CROWN with adjustable lower bounds.}
  \label{Fig_ReLU}
\end{figure*}
A commonly used incomplete verifier is the linear programming (LP) verifier, which relaxes non-convex ReLU constraints with linear constraints and transforms \eqref{Eq_NN_Optimization} into a LP problem. For $\text{ReLU}(z_j^{(i)}) := \max(0, z_j^{(i)})$ and its intermediate layer bounds $l_j^{(i)} \leq z_j^{(i)} \leq u_j^{(i)}$, each ReLU can be relaxed by the “triangle” relaxation: (1) if $l_j^{(i)} \geq 0$, $\hat{z}_j^{(i)} = z_j^{(i)}$ (Fig. \ref{Fig_ReLU}-a); (2) if $l_j^{(i)} \leq 0$, $\hat{z}_j^{(i)} = 0$ (Fig. \ref{Fig_ReLU}-b); (3) if $l_j^{(i)} \leq 0 \leq u_j^{(i)}$, three linear bounds are used: $\hat{z}_j^{(i)} \geq 0$, $\hat{z}_j^{(i)} \geq z_j^{(i)}$, and $\hat{z}_j^{(i)} \leq \frac{u_j^{(i)}}{u_j^{(i)} - l_j^{(i)}} \left( z_j^{(i)} - l_j^{(i)} \right)$ (Fig. \ref{Fig_ReLU}-c). However, LP verifiers use fixed intermediate bounds and cannot use the joint optimization of intermediate layer bounds to tighten relaxation.

\subsection{CROWN and \texorpdfstring{$\alpha$-CROWN}{alpha-CROWN}}\label{sec3d}
Another cheaper way to find $\underline f$ is CROWN, which backward propagates a linear bound from $f$ to each intermediate layer, and finally to the input $\bm x$. CROWN uses a linear upper bound $\hat{z}_j^{(i)} \leq \frac{u_j^{(i)}}{u_j^{(i)} - l_j^{(i)}} \left( z_j^{(i)} - l_j^{(i)} \right)$ and a linear lower bound $\hat{z}_j^{(i)} \geq \alpha_j^{(i)} z_j^{(i)} ~ (0 \leq \alpha_j^{(i)} \leq 1)$ to relax non-convex ReLU, as shown in Fig. \ref{Fig_ReLU}-d. The ReLU relaxation in CROWN is
\begin{equation}
\label{Eq_ReLU}
    \bm{W}^\top \text{ReLU}(\bm z) \geq  \bm{W}^\top \bm{D} \bm z + \bm b'
\end{equation}
where $\bm{D}$ is a diagonal matrix: $\bm{D}_{j,j} = 1$ (if $l_j \geq \bm 0$), $\bm{D}_{j,j} = 0$ (if $u_j \leq 0$), $\bm{D}_{j,j} = \alpha_j$ (if $l_j < 0 < u_j$ and $W_j \geq 0$), and $\bm{D}_{j,j} = \frac{u_j}{u_j - l_j}$ (if $l_j < 0 < u_j$ and $W_j < 0$); $\bm{b}' = \bm{W}^\top \bm{b}$ and $\bm{b}_j = 0$ (if $l_j > 0$ or $u_j \leq 0$ or ($l_j < 0 < u_j$ and $W_j \geq 0$)), and $\bm{b}_j = -\frac{u_j l_j}{u_j - l_j}$ (if $l_j < 0 < u_j$ and $W_j < 0$). The CROWN can be formulated as \cite{zhang2018efficient}:
\begin{equation}
\label{Eq_CROWN}
    f^*_{\text{CROWN}} = \min_{\bm x \in \mathcal{C}} f(\bm x) \geq \min_{\bm x \in \mathcal{C}} \bm{a}_{\text{CROWN}}^\top \bm x + c_{\text{CROWN}}
\end{equation}
where $\bm{a}_{\text{CROWN}}$ and $c_{\text{CROWN}}$ can be calculated using $\bm{W}^{(i)}$, $\bm{b}^{(i)}$, $\bm{l}^{(i)}$ and $\bm{u}^{(i)}$ in polynomial time. \eqref{Eq_CROWN} can be solved using Hölder’s inequality and accelerated by GPUs and TPUs \cite{xu2020automatic}. Based on CROWN, $\alpha$-CROWN further uses gradient ascent to optimize $\alpha$ and tighten the lower bound, as shown in Fig. \ref{Fig_ReLU}-e. The $\alpha$-CROWN can be formulated as \cite{xu2020fast}:
\begin{equation}
\label{Eq_alpha_CROWN}
    f^*_{\alpha-\text{CROWN}} = \min_{\bm x \in \mathcal{C}} f(\bm x) \geq \max_{0 \leq \bm \alpha \leq 1} \min_{\bm x \in \mathcal{C}} f_{\text{CROWN}}(\bm x; \bm \alpha)
\end{equation}

\subsection{\texorpdfstring{$\beta$-CROWN}{beta-CROWN}}\label{sec3e}
\begin{figure}[htb]
  \centering
  \includegraphics[width=8.8cm]{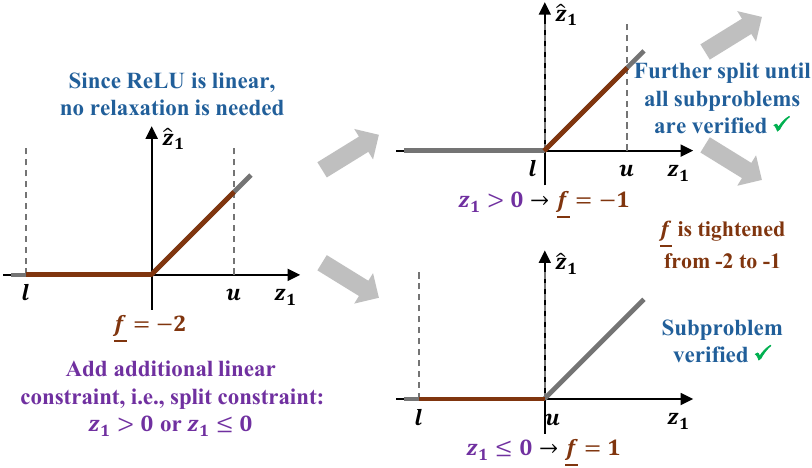}
  \caption{BaB-based NN splitting and verification.}
  \label{Fig_BaB}
\end{figure}

Branch and bound (BaB) method is widely used in complete verifiers by dividing the $\mathcal{C}$ into two linear subdomains $\mathcal{C}_1 = \{ x \in \mathcal{C}, z_j^{(i)} \geq 0 \}$ and $\mathcal{C}_1 = \{ x \in \mathcal{C}, z_j^{(i)} < 0 \}$. Then incomplete verifiers can be used to relax and estimate the lower bound of each subdomain. Since ReLU is linear, no relaxation is needed. If the lower bound for $\mathcal{C}_i$ is greater than 0, i.e., $\underline{f}_{\mathcal{C}_i} > 0$, $\mathcal{C}_i$ is verified; otherwise, $\mathcal{C}_i$ needs to further branch other unstable ReLUs until it is verified, as shown in Fig. \ref{Fig_BaB}

The neuron splitting-based NN verification can be formulated as an optimization problem:
\begin{equation}
\label{Eq_beta_neuron_splitting}
    \min_{\bm x \in \mathcal{C}, \bm z \in \mathcal{Z}} f(\bm x) = \min_{\bm x \in \mathcal{C}, \bm z \in \mathcal{Z}} \bm{W}^{(L)} \bm{\hat{z}}^{(L-1)} + \bm{b}^{(L)}
\end{equation}

Since $\bm{\hat{z}}^{(L-1)} = \text{ReLU}(\bm z^{(L-1)})$, use \eqref{Eq_ReLU} to relax ReLU neuron at layer $L-1$ to obtain:
\begin{equation}
\label{Eq_beta_CROWN}
    \min_{\bm x \in \mathcal{C}, \bm z \in \mathcal{Z}} f(\bm x) \geq \min_{\bm x \in \mathcal{C}, \bm z \in \mathcal{Z}} \bm{W}^{(L)} \bm{D}^{(L-1)} \bm z^{(L-1)} + c
\end{equation}
where $c$ denotes all constant terms. Then, the Lagrange function is used to eliminate the constraint $\bm z^{(L-1)} \in \mathcal{Z}^{(L-1)}$ by adding $\bm{\beta}^{(L-1)^{\top}} \bm{S}^{(L-1)} \bm z^{(L-1)}$, where $\bm\beta$ denotes the Lagrange multiplier; $\bm S$ is a diagonal matrix that encodes the split state of the neurons: $\bm{S}^{(i)}_{j,j} = -1$ (if split $z^{(i)}_j \geq 0$), $\bm{S}^{(i)}_{j,j} = +1$ (if split $z^{(i)}_j < 0$), and $\bm{S}^{(i)}_{j,j} = 0$ (if no split for $z^{(i)}_j$):
\begin{equation}
\label{Eq_beta_Lagrange}
\begin{split}
    \min_{\bm x \in \mathcal{C}, \bm z \in \mathcal{Z}} f(\bm x) \geq & \min_{\substack{\bm x \in \mathcal{C} \\
    \tilde{\bm z}^{(L-2)} \in \tilde{\mathcal{Z}}^{(L-2)}}} \max_{\bm{\beta}^{(L-1)} \geq \bm 0} \bm{W}^{(L)} \bm{D}^{(L-1)} \bm z^{(L-1)} \\ & + \bm{\beta}^{(L-1)^{\top}} \bm{S}^{(L-1)} \bm z^{(L-1)} + c \\
    \geq & \max_{\bm{\beta}^{(L-1)} \geq \bm 0} \min_{\substack{\bm x \in \mathcal{C} \\ \tilde{\bm z}^{(L-2)} \in \tilde{\mathcal{Z}}^{(L-2)}}} \left(\bm{W}^{(L)} \bm{D}^{(L-1)} \right. \\ & + \left. \bm{\beta}^{(L-1)^{\top}} \bm{S}^{(L-1)}\right) \bm z^{(L-1)} + c
\end{split}
\end{equation}
where the second inequality is based on the weak duality; $\tilde{\mathcal{Z}}^{(i)}:=\mathcal{Z}^{(1)} \cap \cdots \cap \mathcal{Z}^{(i)}$; $\tilde{\bm z}^{(i)} := \{\bm z^{(1)}, \bm z^{(2)}, \dots, \bm z^{(i)}\}$. Then $\bm z^{(L-1)}$ is substituted by $\bm{W}^{(L-1)} \bm{\hat{z}}^{(L-2)}$ for next layer:
\begin{equation}
\label{Eq_beta_z_L_1}
\begin{split}
    \min_{\bm x \in \mathcal{C}, \bm z \in \mathcal{Z}} f(\bm x) \geq & \max_{\bm\beta^{(L-1)} \geq \bm 0} \min_{\substack{\bm x \in \mathcal{C} \\\tilde{\bm z}^{(L-2)} \in \tilde{\mathcal{Z}}^{(L-2)}}} \left( \bm{W}^{(L)} \bm{D}^{(L-1)} \right. \\& \left.+ \bm\beta^{(L-1)\top} \bm{S}^{(L-1)} \right) \bm{W}^{(L-1)} \bm{\hat{z}}^{(L-2)} + c
\end{split}
\end{equation}

Finally, by recursively back-substituting the linear constraints of the NN layer by layer and swapping max and min at each step, \eqref{Eq_beta_z_L_1} becomes:
\begin{equation}
\label{Eq_beta_CROWN_Lemma}
    \min_{\bm x \in \mathcal{C}, z \in \mathcal{Z}} f(\bm x) \geq \max_{\bm \beta \geq \bm 0} \min_{\bm x \in \mathcal{C}} \left( \bm{a} + \bm{P}\bm \beta \right)^\top \bm x + \bm{q}^\top \bm \beta + c
\end{equation}
where $\bm\beta := \left[ \bm\beta^{(1)^\top}, \bm\beta^{(2)^\top}, \dots, \bm\beta^{(L-1)^\top} \right]^\top$; $\bm{a}$, $\bm{P}$, $\bm{q}$, and $c$ are functions of $\bm{W}^{(i)}$, $\bm{b}^{(i)}$, $\bm{l}^{(i)}$ and $\bm{u}^{(i)}$.
In the $\ell_p$ norm ball $\mathcal{C} = \left\{\bm x \ \middle| \ \|\bm x - \bm{\widetilde{x}}\|_p \leq \bm \epsilon \right\}$, the inner minimization has a closed solution:
\begin{equation}
\label{Eq_beta_inner_minimization}
\begin{split}
    \min_{\bm x \in \mathcal{C}, \bm z \in \mathcal{Z}} f(\bm x) \geq & \max_{\bm\beta \geq \bm 0} -\| \bm{a} + \bm{P} \bm\beta \|_{q} \bm\epsilon + \left( \bm{P}^\top \bm{\widetilde{x}} + \bm{q} \right)^\top \bm\beta \\ & + \bm{a}^\top \bm{\widetilde{x}} + c := \max_{\bm\beta \geq \bm 0} g(\bm \beta)    
\end{split}
\end{equation}
where $\frac{1}{p} + \frac{1}{q} = 1$. The maximization is concave in $\bm\beta$ (when $q \geq 1$) and can be optimized using projected (super) gradient ascent. Then, the slope $\bm\alpha = \{\bm\alpha^{(1)}, \dots, \bm\alpha^{(L-1)}\}$ of the ReLU lower bound is optimized to further tighten the bound, and \eqref{Eq_beta_inner_minimization} can be rewritten as:
\begin{equation}
\label{Eq_beta_CROWN_Final}
    \min_{\bm x \in \mathcal{C}, \bm z \in \mathcal{Z}} f(\bm x) \geq \max_{\bm 0 < \bm\alpha \leq \bm 1, \bm\beta \geq \bm 0} g(\bm\alpha, \bm\beta)
\end{equation}

\begin{algorithm}[tb]
\label{Algorithm_NN_Train}
\SetAlgoNoLine
\caption{DBN Training for Transient Stability Classification (DBN-C) and Estimation (DBN-E).}
\KwIn{Historical dataset of $\bm{x}$}
\KwOut{DBN-C and DBN-E}
Set fault contingency\\
Run TDS to calculate TSI\\
Data preprocessing and normalization\\
Build DBN-C and DBN-E in PyTorch, where DBN-C outputs the 0/1 classification labels for TSI, and DBN-E outputs the TSI\\
Set the number of layers, the number of neurons per layer, the activation function ReLU, and other hyperparameters\\
Set the cross entropy loss for DBN-C and the mean squared error loss for DBN-E, respectively\\
Train DBN-C and DBN-E
\end{algorithm}
\begin{algorithm}[tb]
\label{Algorithm_Verification}
\SetAlgoNoLine
\caption{$\alpha, \beta$-CROWN-Based Transient Stability Robustness
Verification.}
\KwIn{$\bm{\widetilde{x}}$, DBN-C}
\KwOut{Verification results: Safety = safe or unsafe or unknow}
Set $\mathcal{C}$ according to the possible perturbation range of $\bm{P}_{\text{IBR}}$, $\bm{P}_{\text{SG}}$, $\bm{P}_{d}$, and $\bm{Q}_{d}$\\
Set \textit{Safety = unknown}\\
Perform PGF attack according to \eqref{Eq_PGD_Attack}\\
\If{PGD attack == True}
    {\textbf{Return}: Safety = unsafe}
Perform $\alpha$-CROWN according to \eqref{Eq_alpha_CROWN}\\
\If{Safety == unknown \textbf{and} $\alpha$-CROWN == True}
    {\textbf{Return}: Safety = safe}
Perform $\beta$-CROWN according to \eqref{Eq_beta_CROWN_Final}\\
\If{Safety == unknown \textbf{and} $\beta$-CROWN == True}
    {\textbf{Return}: Safety = safe}
\textbf{Return}: Safety = unknown
\end{algorithm}
\begin{algorithm}[tb]
\label{Algorithm_TSPC}
\SetAlgoNoLine
\caption{$\alpha, \beta$-CROWN-Based Transient Stability Preventive Control.}
\KwIn{$\bm{\overline{P}}_{\text{IBR}}$, $\bm{\widetilde{P}}_{d}$, $\bm{\widetilde{Q}}_{d}$, DBN-E}
\KwOut{Safety-verified preventive control strategies $\bm{\widetilde{x}}$}
Initialize the $\bm{\widetilde{P}}_{\text{IBR}}$ and $\bm{\widetilde{P}}_{\text{SG}}$\\
Set $\lambda = 0$, $\lambda_{\text{left}} = 0$, $\lambda_{\text{right}} = 90$, $\Delta \lambda = 2$, and $\zeta = 1$\\
\While{$|\Delta \lambda| \geq \zeta$}
{
    Perform TSC-OPF using PDIPM according to the objective function \eqref{Eq_objective}, and constraints \eqref{Eq_PF_P} - \eqref{Eq_S_IBR} and \eqref{Eq_DBN_Estimation} \\
    Verify $\bm{\widetilde{x}}$ according to Algorithm \ref{Algorithm_Verification}\\
    \If{Converge == True \textbf{and} Safety != safe}
        {$\Delta \lambda = \lambda_{\text{right}} - \lambda$, $\lambda_{\text{left}} = \lambda$, $\lambda = \lambda_{\text{left}} + \Delta \lambda / 2$}
    \ElseIf{Converge == False \textbf{and} Safety != safe}
        {No feasible solution. Reduce the range of $\mathcal{C}$ or implement other control measures, such as load shedding.}
    \Else
        {$\Delta \lambda = \lambda - \lambda_{\text{left}}$, $\lambda_{\text{right}} = \lambda$, $\lambda = \lambda_{\text{left}} + \Delta \lambda / 2$}
}
\end{algorithm}

\begin{figure*}[htb]
  \centering
  \includegraphics[width=16cm]{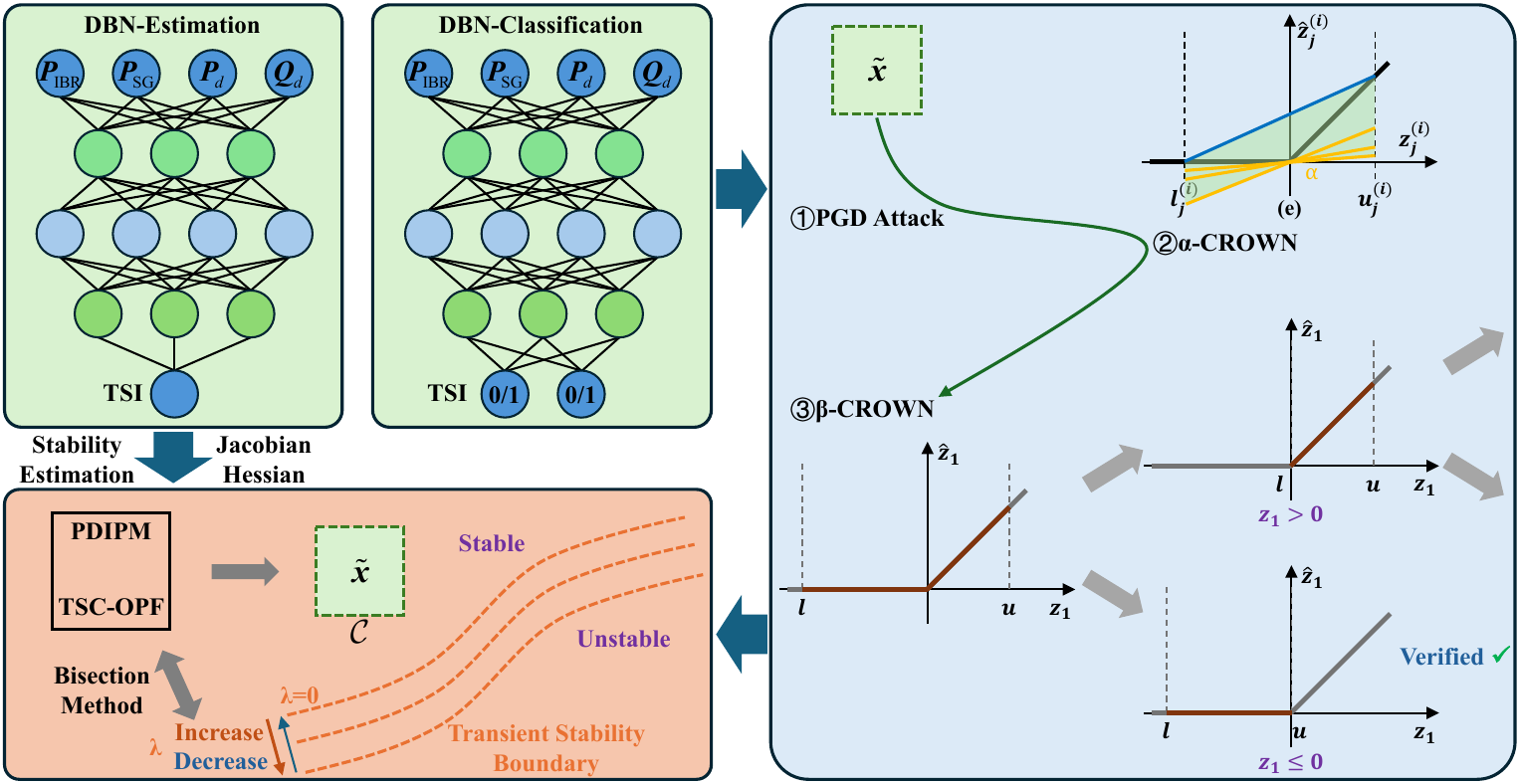}
  \caption{Proposed $\alpha, \beta$-CROWN-based transient stability preventive control framework.}
  \label{Fig_Framework}
\end{figure*}

\subsection{\texorpdfstring{$\alpha, \beta$-CROWN-Based Transient Stability Preventive Control}{alpha, beta-CROWN-Based Transient Stability Preventive Control}}\label{sec3f}
The proposed $\alpha, \beta$-CROWN-based transient stability preventive control is divided into three parts:

1) Two different DBN models are trained, one for classification and one for estimation. This is because $\alpha, \beta$-CROWN requires a classification model to verify if the lower bound misclassifies into another category, while TSC-OPF requires a DBN-based transient stability estimation surrogate model to replace TDS and provides gradients of TSI with respect to control strategy $\bm{\widetilde{x}}$ to guide the iteration process. The detailed steps of DBN training are shown in Algorithm \ref{Algorithm_NN_Train}.

2) Verify the preventive control strategy $\bm{\widetilde{x}}$ using $\alpha, \beta$-CROWN. Due to the increasing verification time with PGD attacks, $\alpha$-CROWN, and $\beta$-CROWN, execute them sequentially. If a verification result is obtained, return directed; otherwise, proceed to the next verification method. The results of NN verification are categorized into three types: safe (no misclassification exists within $\mathcal{C}$, unsafe (there are misclassified points within $\mathcal{C}$), and unknown (the robustness of $\mathcal{C}$ cannot be verified). The verification steps are shown in Algorithm \ref{Algorithm_Verification}. 

3) Execute TSC-OPF, where the DBN is embedded as a surrogate model for TDS to accelerate the computation. This nonlinear and non-convex problem is solved using PDIPM, where the DBN leverages PyTorch’s automatic differentiation function to calculate the Jacobian and Hessian matrices of TSI with respect to $\bm{\widetilde{x}}$, guiding the PDIPM iteration to obtain the final preventive control strategy $\bm{\widetilde{x}}$. Afterward, the robustness of $\bm{\widetilde{x}}$ is verified using the $\alpha, \beta$-CROWN described in the Algorithm \ref{Algorithm_Verification}. Since $\lambda$ determines the safety margin of the transient stability constraint, different values of $\lambda$ can be set to control the safety margin of $\bm{\widetilde{x}}$. The value of $\lambda$ can be computed using the bisection method: if the verification result is safe, reduce $\lambda$ (with a minimum of 0); otherwise, increase $\lambda$ until the verification result is safe, resulting in the final safety-verified $\bm{\widetilde{x}}$. When TSC-OPF does not converge and the NN verification fails, the problem is unsolvable. The details are shown in Algorithm \ref{Algorithm_TSPC}.

The whole framework of the proposed method is shown in Fig. \ref{Fig_Framework}.

\section{Numerical Results}\label{sec4}
\begin{figure}[htb]
  \centering
  \includegraphics[width=8.8cm]{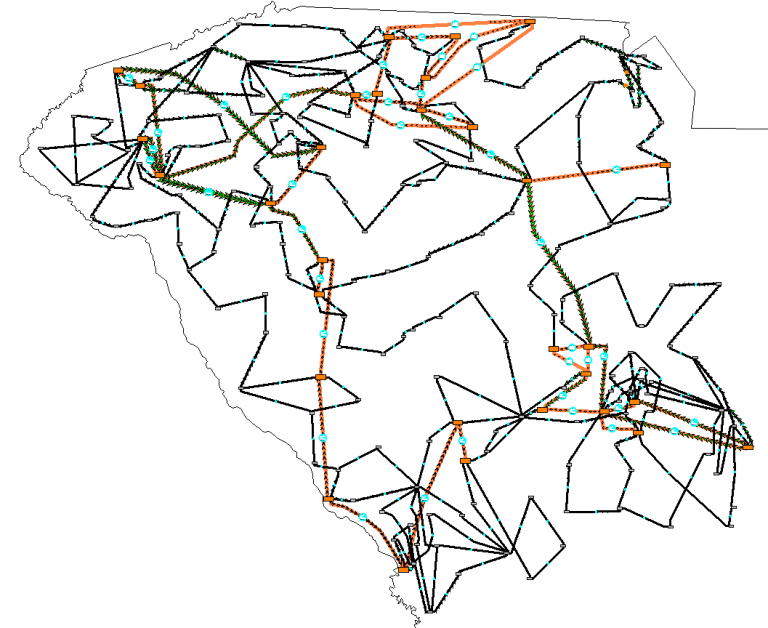}
  \caption{One-line diagram of the South Carolina 500-bus system.}
  \label{Fig_500}
\end{figure}

Numerical results are tested on the South Carolina 500-bus system, which serves 21 counties and approximately 2.6 million people, with 206 loads supplied by 9 IBRs and 51 SGs, as shown in Fig. \ref{Fig_500} \cite{xu2017creation}.

\subsection{\texorpdfstring{$\alpha, \beta$-CROWN-Based Transient Stability Robustness Verification}{alpha, beta-CROWN-Based Transient Stability Robustness Verification}}\label{sec4a}
According to Algorithm \ref{Algorithm_NN_Train}, two 5-layer DBN models are built in PyTorch for transient stability classification (DBN-C) and estimation (DBN-E), respectively. The classification accuracy of DBN-C is 99.15\%, and the mean absolute error of estimation by DBN-E is 1.86, demonstrating excellent accuracy. The detailed training and comparison results of DBN are shown in \cite{su2021deep}. The classification or estimation time of DBN is less than 0.01s, and meets the requirement of online application and subsequent robustness verification and preventive control.

To demonstrate the transient stability robustness verification performance of $\alpha, \beta$-CROWN, 1,000 random samples are assumed as $\bm{\widetilde{x}}$, and their safety is verified using $\alpha, \beta$-CROWN. The maximum perturbation range for IBRs, SGs, and loads is set to 5\% to construct $\mathcal{C}$. Then, $\alpha, \beta$-CROWN is used to test the probability that all points in $\mathcal{C}$ maintain the same classification as $\bm{\widetilde{x}}$. The verification process is shown in Algorithm \ref{Algorithm_Verification}, and the verification results are shown in Table \ref{Table_Verify}.
\begin{table}[htb]
\caption{Statistical Probabilities of NN Robustness Verification Results for Random Scenarios}
\label{Table_Verify}
\centering
\begin{tabular}{ccccc}
    \toprule 
    unsafe-PGD & safe-incomplete & safe-complete & unknown \\
    \midrule 
    63.4\% & 29.7\% & 6.5\% & 0.4\%\\
    \bottomrule 
\end{tabular}
\end{table}
In Table \ref{Table_Verify}, unsafe-PGD indicates that the classification result changed due to the PGD attack; safe-incomplete means the verification result is classified as safe by $\alpha$-CROWN, while safe-complete means it is classified as safe by $\beta$-CROWN; unknown indicates that the verification result is still unknown after $\alpha, \beta$-CROWN. Since $\alpha, \beta$-CROWN progressively verifies robustness through PGD attacks, $\alpha$-CROWN, and $\beta$-CROWN, 63.4\% of the samples are verified as unsafe by PGD attacks, followed by 29.7\% of the samples verified as safe by $\alpha$-CROWN. Finally, 6.9\% of the samples needed verification by $\beta$-CROWN, among which 6.5\% are found to be safe, and 0.4\% are unsafe. From the verification results, it can be observed that $\alpha, \beta$-CROWN can obtain the robustness verification result for the preventive control strategy $\bm{\widetilde{x}}$ in 99.6\% of cases, and this result is complete. In addition, compared to the traditional approach based solely on PGD attacks, the verification success rate improved from 63.4\% to 99.6\%.

\begin{figure}[htb]
  \centering
  \includegraphics[width=8.8cm]{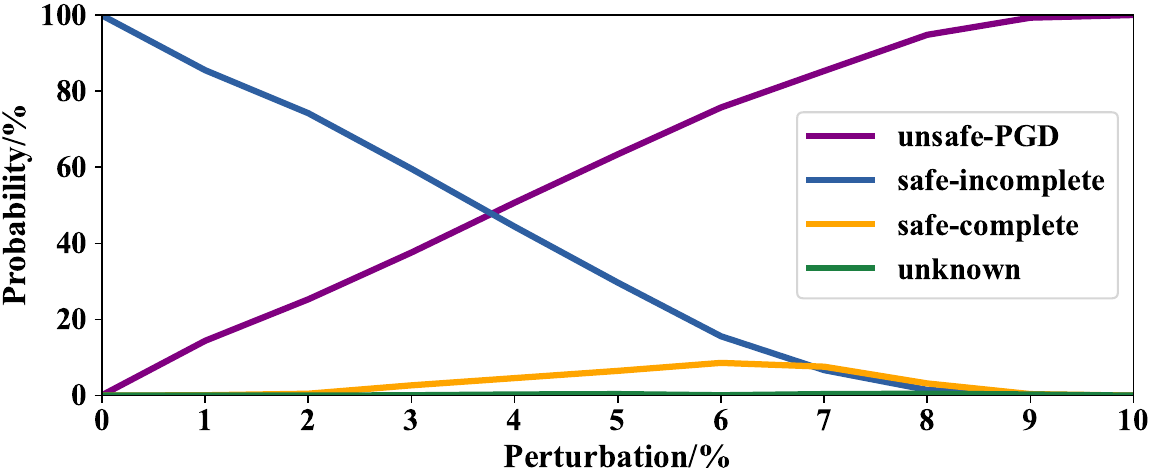}
  \caption{Comparison of NN verification results with different perturbations.}
  \label{Fig_Perturbation}
\end{figure}

To further verify the effects of different perturbations, the previously used 1,000 samples are tested with perturbations varying from 0\% to 10\%. The metrics tested include unsafe-PGD, safe-incomplete, safe-complete, and unknown, as shown in Fig. \ref{Fig_Perturbation}.
From \ref{Fig_Perturbation}, it can be observed that when the perturbation is very small, the margin between $\mathcal{C}$ and the transient stability boundary is large. In this case, PGD attacks are almost unable to determine whether the $\bm{\widetilde{x}}$ is unsafe, while $\alpha$-CROWN can almost completely verify the robustness. As the perturbation increases, the margin between $\mathcal{C}$ and the transient stability boundary decreases, allowing more $\bm{\widetilde{x}}$ to be verified by PGD attacks. However, due to the looser lower bounds, incomplete verification via $\alpha$-CROWN may be unable to determine the robustness of some samples, necessitating complete NN verification through $\beta$-CROWN. As the perturbation grows closer to 10\%, $\mathcal{C}$ for most samples $\bm{\widetilde{x}}$ crosses the transient stability boundary, allowing PGD attacks to verify the robustness. In addition, the probability of samples being verified as unknown is very low, consistently below 0.5\%.

\begin{figure}[htb]
  \centering
  \includegraphics[width=8.8cm]{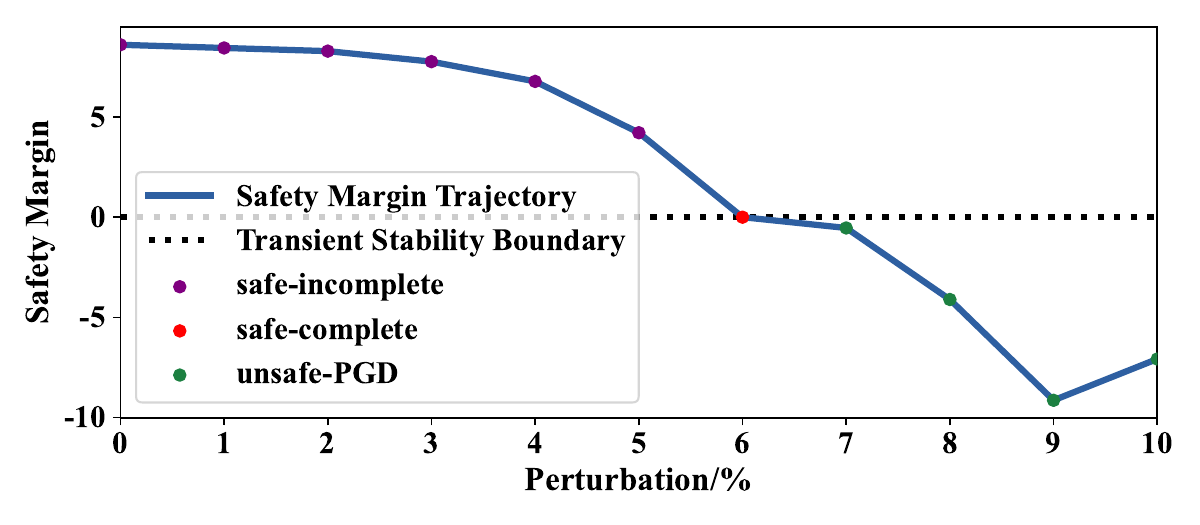}
  \caption{Calculation of maximum possible perturbation and safety margin.}
  \label{Fig_Safety_Margin}
\end{figure}

$\alpha, \beta$-CROWN can also be used to determine the maximum possible perturbation and safety margin. For example, for a specific $\bm{\widetilde{x}}$, start with zero perturbation and gradually increase it up to 10\%, verifying robustness at each step, as shown in Fig. \ref{Fig_Safety_Margin}. 
It is important to point out that safe-incomplete, safe-complete, and unsafe-PGD correspond to different meanings of the safety margin. Safe-incomplete refers to the distance from the $\alpha$-CROWN lower bound to 0, while unsafe-PGD corresponds to the difference between the values of two classifications. Safe-complete uses $\beta$-CROWN based on BaB, which further splits unstable ReLUs when the result is inconclusive, until the lower bound is greater than 0. Therefore, its final margin is a small positive value, but in practice, if all ReLUs are split, the margin could be further increased. Therefore, in Fig. \ref{Fig_Safety_Margin}, within the 0–5\% perturbation range, as the perturbation increases, the safety margin between the $\alpha$-CROWN lower bound and the transient stability boundary gradually decreases, from 8.606 to 4.216. At 6\% perturbation, the safety margin is $1 \times 10^{-7}$, meaning that splitting stopped once it became greater than 0. In the 7–10\% perturbation range, the safety margin is always less than 0, indicating that the value corresponding to the classification of $\bm{\widetilde{x}}$ is less than that of the other class. Since a 6\% perturbation is classified as safe-complete and a 7\% perturbation is classified as unsafe-PGD, the maximum possible perturbation for this particular $\bm{\widetilde{x}}$ lies between 6\% and 7\%. If a more precise value is desired, further refinement between 6\% and 7\% can be performed. Additionally, the bisection method could be used to converge more quickly to this value. Since different verifications are completely decoupled, multiple servers can be used in parallel to determine this value more efficiently.

The average verification time for PGD attacks is only 0.00795s, while the average verification time for $\alpha$-CROWN and $\beta$-CROWN is 0.495s. The overall average verification time for $\alpha, \beta$-CROWN is 0.196s. From the verification times, it is evident that executing the PGD attack first can significantly save time. However, the verification times of all methods are very short, fully meeting the requirements of preventive control.

\subsection{\texorpdfstring{$\alpha, \beta$-CROWN-Based Transient Stability Preventive Control}{alpha, beta-CROWN-Based Transient Stability Preventive Control}}\label{sec4b}
\begin{figure}[htb]
  \centering
  \includegraphics[width=8.8cm]{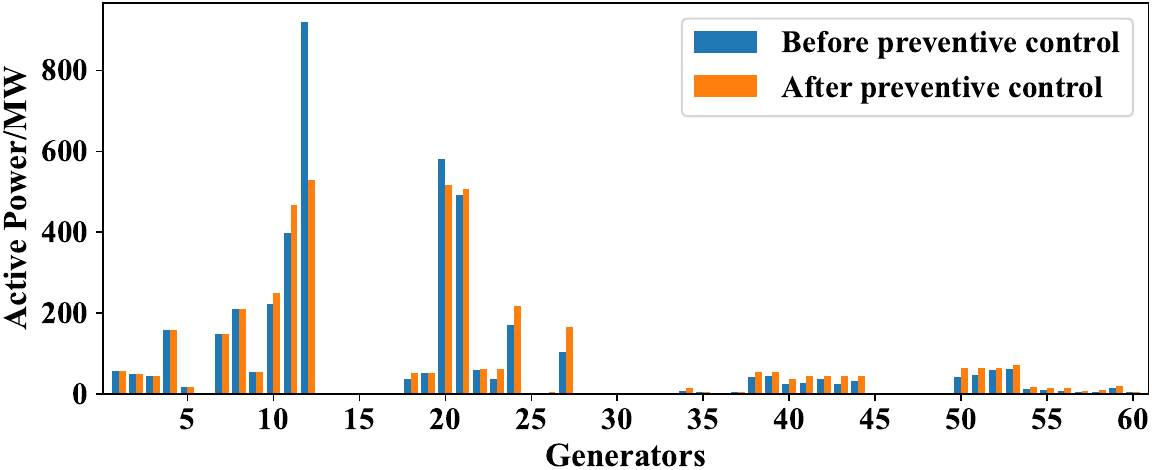}
  \caption{Comparison of active power generation of IBRs and SGs before and after preventive control without NN robustness verification.}
  \label{Fig_Gen_Dispatch_without_Verification}
\end{figure}

\begin{figure}[htb]
  \centering
  \includegraphics[width=4.35cm]{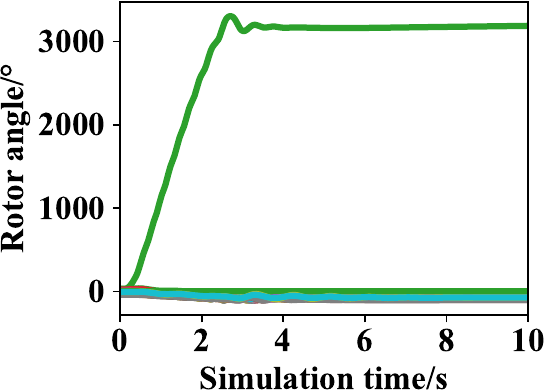}
  \includegraphics[width=4.35cm]{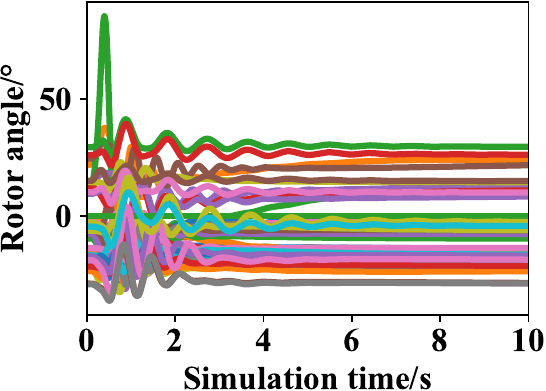}
  \caption{Rotor angle trajectories before and after preventive control without robustness verification.}
  \label{Fig_Angle_without}
\end{figure}

\begin{figure}[htb]
  \centering
  \includegraphics[width=4.35cm]{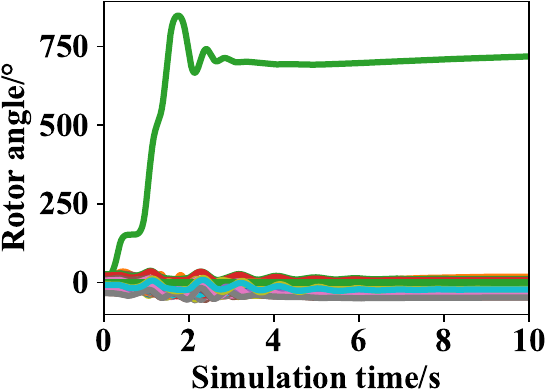}
  \includegraphics[width=4.35cm]{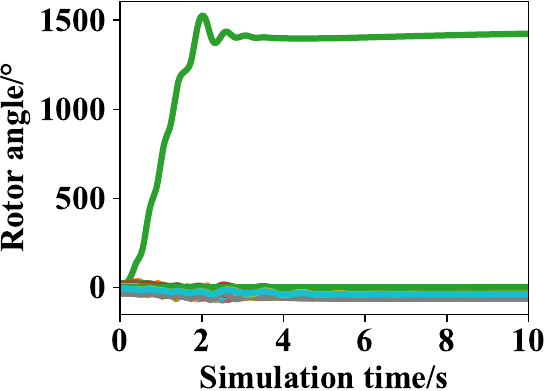}
  \caption{Rotor angle trajectories of $\bm x_{\text{adv}}$ without robustness verification.}
  \label{Fig_Angle_without_noise}
\end{figure}

A TSC-OPF without NN verification is performed to find the optimal $\lambda$. The preventive control strategy $\bm{\widetilde{x}}$ for the active power generation of IBRs and SGs, compared to the $\bm{x}$ without preventive control, is shown in Fig. \ref{Fig_Gen_Dispatch_without_Verification}.
In Fig. \ref{Fig_Gen_Dispatch_without_Verification}, the first 9 generators are IBRs, and the remaining 51 generators are SGs. It can be observed that all IBRs maintain the same active power output as $\bm{\overline{P}}_{\text{IBR}}$, indicating no RES curtailment, thus minimizing cost. Meanwhile, the active power generations of all SGs are dispatched to minimize fuel costs. The total cost is \$19,035.6. The rotor angle trajectories before and after preventive control are shown in Fig. \ref{Fig_Angle_without}.
From Fig. \ref{Fig_Angle_without}, it can be observed that, after preventive control, the control strategy $\bm{\widetilde{x}}$ changes the power angle trajectory from divergent to convergent, transitioning from transient instability to stability. However, when considering perturbations of $\bm{\widetilde{x}}$, there may be unstable points within $\mathcal{C}$. To verify this, set the maximum perturbation range for IBRs to 20\%, for SGs to 8\%, and for loads to 10\%, to construct $\mathcal{C}$. It is important to note that in Fig. \ref{Fig_Perturbation}, when the perturbation reaches 10\%, almost all samples are classified as unsafe-PGD. However, a larger perturbation, such as 20\% for IBRs, can be set here because the safety and robustness of the preventive control strategies from TSC-OPF have been significantly improved overall. Then, MCS is used to randomly sample points from $\mathcal{C}$ and verify their power angle trajectories. After hundreds of MCS samples, two adversarial samples $\bm x_{\text{adv}}$ are found, as shown in Fig. \ref{Fig_Angle_without_noise}, indicating the potential risks of directly applying the preventive control strategy $\bm{\widetilde{x}}$.

\begin{figure}[htb]
  \centering
  \includegraphics[width=8.8cm]{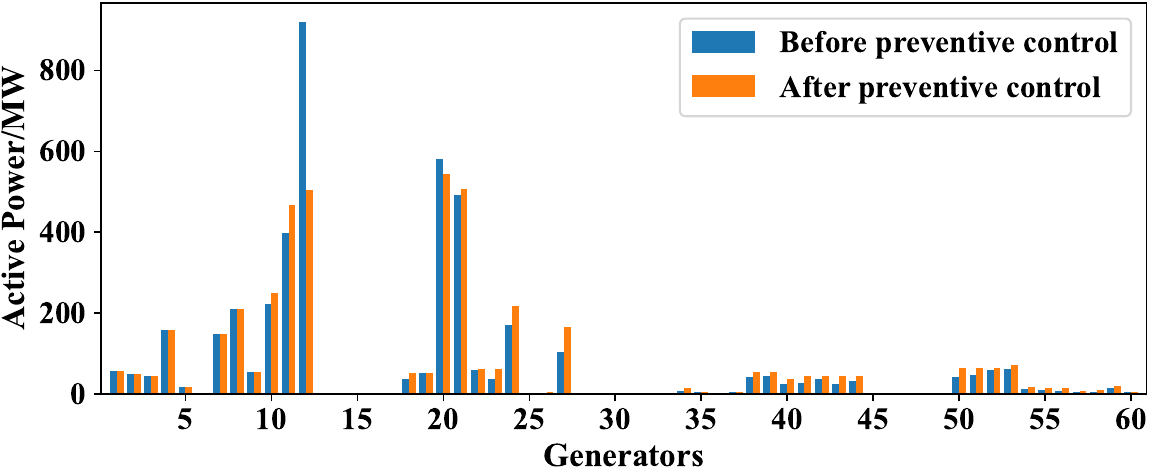}
  \caption{Comparison of active power generation of IBRs and SGs before and after preventive control with NN robustness verification.}
  \label{Fig_Gen_Dispatch_with_Verification}
\end{figure}

\begin{figure}[htb]
  \centering
  \includegraphics[width=4.35cm]{Fig_angle_before.pdf}
  \includegraphics[width=4.35cm]{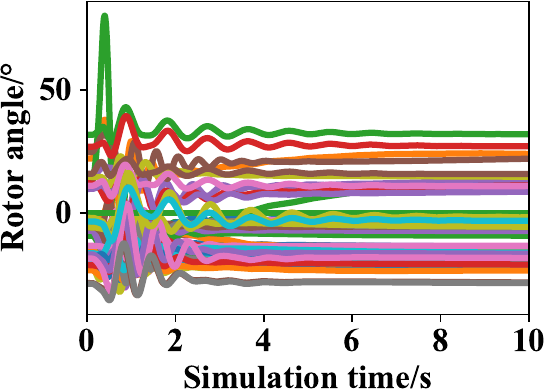}
  \caption{Rotor angle trajectories before and after preventive control with robustness verification.}
  \label{Fig_Angle_with}
\end{figure}

\begin{figure}[htb]
  \centering
  \includegraphics[width=4.35cm]{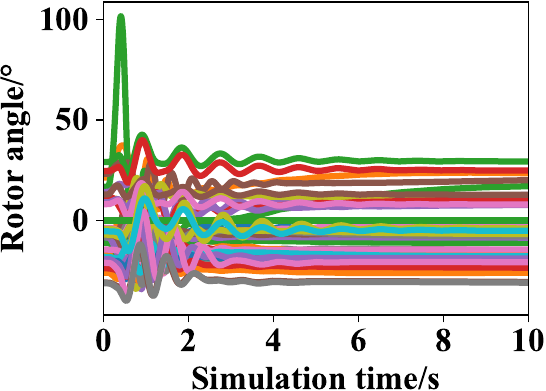}
  \includegraphics[width=4.35cm]{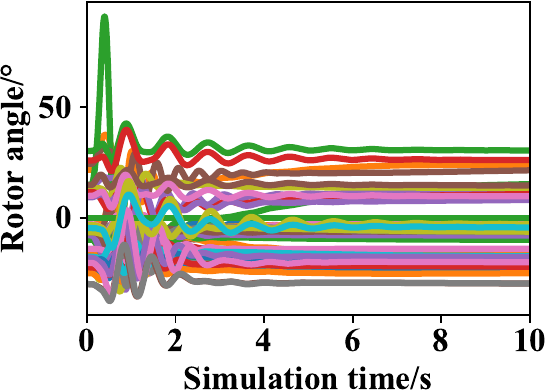}
  \caption{Rotor angle trajectories of points in $\mathcal{C}$ with robustness verification.}
  \label{Fig_Angle_with_noise}
\end{figure}

Next, using the same $\mathcal{C}$, TSC-OPF is performed according to Algorithm \ref{Algorithm_TSPC}, and the bisection method is used to find the optimal $\lambda$ when robustness verification fails. The iterative process is shown in Table \ref{Table_Bisection_Method}. 
\begin{table}[htb]
\caption{Statistical Probabilities of NN Robustness Verification Results for Random Scenarios}
\label{Table_Bisection_Method}
\centering
\begin{tabular}{cccccc}
    \toprule 
    Ite. & $\lambda$ & Converge & Verification & Cost/\$ & TSI \\
    \midrule 
    1 & 0.000000 &  True &    unsafe-PGD & 19035.6 & 61.26032 \\
    2 & 45.00000 &  True &    unsafe-PGD & 19035.6 & 61.26032 \\
    3 & 67.50000 & False & safe-complete & 20431.9 & 65.75372 \\
    4 & 56.25000 &  True &    unsafe-PGD & 19035.6 & 61.26032 \\
    5 & 61.87500 &  True &    unsafe-PGD & 19037.5 & 61.87499 \\
    6 & 64.68750 &  True & safe-complete & 19102.4 & 64.68749 \\
    7 & 63.28125 &  True & safe-complete & 19057.8 & 63.28125 \\
    \bottomrule 
\end{tabular}
\end{table}
From Table \ref{Table_Bisection_Method}, it can be observed that when $\lambda$ is too large, there is a risk of TSC-OPF being diverged. This occurs because the safety margin for transient stability constraints becomes excessively large, resulting in no feasible solution. When $\lambda \leq 61.875$, as in iterations 1, 2, 4, and 5, the safety margin is insufficient, and the PGD attack succeeds, indicating that these $\bm{\widetilde{x}}$ are unsafe. When $\lambda = 67.5$, as in iteration 3, even though TSC-OPF does not converge, the verification result is safe-complete. This occurs because, while the transient stability constraint is satisfied, other constraints within TSC-OPF are not, leading to non-convergence. From the cost and TSI, it can be seen that in some iterations, even though $\lambda$ is different, the cost and TSI remain unchanged, such as for $\lambda = 0$, $45$, and $56.25$ in iterations 1, 2, and 4. This is because $\lambda$ is relatively small ($\lambda < 61.26032$), causing the TSC-OPF to be limited by other constraints rather than the TSI constraint. Finally, after 7 iterations, the bisection method converges with $\lambda = \text{TSI} = 63.28125$, indicating that the TSC-OPF has fully converged to the boundary of the transient stability constraint, with robustness also verified as safe. Compared to not performing NN robustness verification, the cost increased by only \$22.2 (from \$19,035.6 to \$19,057.8) after executing NN robustness verification. At this time, the active power generation of IBRs and SGs, compared to $\bm{x}$ without preventive control, is shown in Fig. \ref{Fig_Gen_Dispatch_with_Verification}.
By comparing Figs. \ref{Fig_Gen_Dispatch_without_Verification} and \ref{Fig_Gen_Dispatch_with_Verification}, we find that the generation of most generators remains identical, except for generators 12 and 25, where 25.53 MW is shifted from generator 12 to generator 25 to satisfy the requirement of robustness verification. The rotor angle trajectories before and after preventive control are shown in Fig. \ref{Fig_Angle_with}.
Undoubtedly, the verified $\bm{\widetilde{x}}$ can ensure that the system maintains transient stability. Finally, to further verify the robustness of points within $\mathcal{C}$, similar to Fig. \ref{Fig_Angle_without_noise}, MCS is used to generate a large number of $\bm{x}$ within $\mathcal{C}$ and verify their power angle trajectories. Since all trajectories are stable, only two power angle trajectories are shown in Fig. \ref{Fig_Angle_with_noise}, demonstrating that $\alpha, \beta$-CROWN is a complete robustness verification method.

\section{Conclusion}\label{sec5}
This paper proposes an NN robustness verification-based power system transient stability preventive control method considering errors from various measurement devices, fluctuations in RESs and loads, as well as cyber-attacks. $\alpha, \beta$-CROWN-informed NN certification is developed to verify the robustness of the preventive control strategy within a maximum possible perturbation range. If the verification is unsafe, the bisection method is used to increase the transient stability boundary, thereby increasing the safety margin. The proposed method can balance system security and economics while being scalable to large-scale systems. Numerical results on a modified western South Carolina 500-bus system demonstrate that the proposed method can efficiently and quickly get the safety-verified preventive control strategy through RES curtailment and SG dispatch with only a slight increase in cost.

\bibliographystyle{IEEEtran}
\bibliography{ref.bib}

\end{document}